\documentclass{article}

\let\shortcite\cite
\usepackage{algorithm}
\usepackage{algorithmic}
\usepackage{pifont}
\usepackage{latexsym}
\usepackage{amsfonts}
\usepackage{amsmath}
\usepackage{amssymb}
\usepackage{yhmath}
\newcommand{\p}{{\rm P}}
\newcommand{\np}{{\rm NP}}

\newcommand{\thetatwo}{\ensuremath{\Theta_2^p}}

\newtheorem{theorem}{Theorem}
\newtheorem{corollary}[theorem]{Corollary}

\newtheorem{lemma}[theorem]{Lemma}
\newcommand\qedblob{\ding{113}}
\def\literalqed{{\ \nolinebreak\hfill\mbox{\qedblob\quad}}}

\newtheorem{example}[theorem]{Example}

\newenvironment{proofs}{\noindent{\bf Proof.}\hspace*{1em}}{\literalqed\smallskip}

\showhyphens{}

\newcommand{\ys}{\ensuremath{{\rm YoungScore}}}

\newcommand{\prob}[3]{

\smallskip

\noindent
  {\bf Name:} #1
  
  \noindent
  {\bf Given:} #2
  
  \noindent
  {\bf Question:} #3
  
  \smallskip}
  
\begin{document}

\title{Election Score Can Be Harder Than Winner}

\author{
Zack Fitzsimmons\\
 Dept.\ of Math.\ and Computer Science\\
 College of the Holy Cross\\
Worcester, MA 01610 \and
  Edith Hemaspaandra\\
  Department of Computer Science\\
  Rochester Institute of Technology \\
  Rochester, NY 14623}%

\date{November 20, 2019} %

\sloppy

\maketitle

\begin{abstract}
Election systems based on scores generally determine the winner by computing the score of each candidate and the winner is the candidate with the best score. It would be natural to expect that computing the winner of an election is at least as hard as computing the score of a candidate. We show that this is not always the case. In particular, we show that for Young elections for dichotomous preferences the winner problem is easy, while determining the score of a candidate is hard. This complexity behavior has not been seen before and is unusual. The easiness of the winner problem for dichotomous Young crucially uses the fact that dichotomous preferences guarantee the transitivity of the majority relation. In addition to dichotomous preferences we also look at single-peaked preferences, the most well-studied domain restriction that guarantees the transitivity of the majority relation. We show that for the three major hard election systems and their natural variants, dichotomous Young is the only case where winner is easy and score is hard. This also solves an open question from Lackner and Peters (AAAI 2017), by providing a polynomial-time algorithm for Dodgson score for single-peaked electorates.
\end{abstract}

\section{Introduction}\label{sec:intro}

Election systems based on scores generally find a winner by computing
the score of each candidate and a candidate is a winner if they have the best
score.
Most election systems have easy score problems and so 
finding a winner in this way is also easy. The three most-commonly
studied election systems where this is not the case are Young~\cite{you:j:extending-condorcet}, Dodgson~\cite{dod:unpubMAYBE-without-embedded-citations:dodgson-voting-system}, and
Kemeny~\cite{kem:j:no-numbers} (and its qualitative variant Slater~\cite{sla:j:slater}). The score problem for each of
these systems is NP-complete and each corresponding winner problem is 
\thetatwo-complete~\cite{bar-tov-tri:j:who-won,rot-spa-vog:j:young,hem-hem-rot:j:dodgson,hem-spa-vog:j:kemeny}.

In general it would seem that determining if a candidate is a winner of an election is
at least as hard
as determining the score of a candidate. We show that this is not always
the case. We show that for Young elections for dichotomous preferences the score
problem is hard and the winner problem is easy. This is a behavior that has not been seen
before in the computational study of voting.
Dichotomous preferences are a very natural way for voters to state their preferences, where they approve of a subset of the candidates and disapprove the others. In many situations voters may not be able to state strict preferences
over all of the candidates, but can state their approval/disapproval for each candidate.
Approval voting is the most well-known election system that uses dichotomous 
preferences as input. Additionally, most voting rules %
have definitions that apply or can be naturally extended to dichotomous preferences, and
can keep many of the desirable social-choice properties they have in the total-order case (e.g., this is the case for Kemeny~\cite{bra-pet:j:borda-mean}).

The behavior of Young for dichotomous preferences is quite unusual and does not occur
for any of the obvious variations of the setting.
When we instead
consider strongYoung elections, in which the score is based on how far removed the
candidate is %
from becoming a Condorcet winner (rather than a weak Condorcet winner as in Young),
the score and winner problems for dichotomous preferences are hard. This is also the case when
the electorate is trichotomous (each voter ranks their most preferred, middle preferred, and least preferred
candidates).
Additionally, natural variants of Dodgson and Kemeny for dichotomous preferences do not exhibit
Young's anomalous behavior: Their score and winner problems are in~\p.

The computational easiness of the winner problem for Young for dichotomous preferences follows from the
fact that when the electorate is dichotomous, the majority relation is transitive. To further explore
this effect, we consider additional preference domains.
We examine how having single-peaked preferences and how having single-crossing preferences affects
the complexity of the winner and score problems for Young, Kemeny, and Dodgson elections.
Single-peaked preferences~\cite{bla:j:rationale-of-group-decision-making} and single-crossing
preferences~\cite{mir:j:single-crossing} are the two most-commonly studied domain restrictions that guarantee
transitivity of the majority relation and they each model the preferences of the voters with respect to
a single polarizing issue. In the case of single-peaked preferences, the axis is a total ordering of the candidates
where candidates on the leftmost/rightmost ends of the axis represent the extremes of the issue. In the case of
single-crossing preferences, the voters can be ordered along an axis (in a political setting we can think of of
the leftmost (rightmost) voters as the most-liberal (most-conservative), and for each choice between two alternatives, the preferences of the voters swaps at most once when moving left-to-right along the axis.
We follow the model from Walsh~\shortcite{wal:c:uncertainty-in-preference-elicitation-aggregation} where the single-peaked or single-crossing axis is given as part of the input. However, each of these can be computed in polynomial time~\cite{bar-tri:j:stable-matching-from-psychological-model,elk-fal-sli:c:decloning}.

We consider the complexity of the score and winner problems for Young, strongYoung, Kemeny, Dodgson, and
weakDodgson elections for
single-peaked and for single-crossing preferences. Except for Dodgson and weakDodgson
score for single-crossing preferences, which remain open, all score and winner problems are in \p.
In proving our results, we solve an open problem from
Lackner and Peters~\shortcite{lac-pet:c:spoc} by showing that computing the Dodgson score of a candidate is in P for single-peaked
preferences.

Our study of the relationship between score and winner is significant
in two ways. The first one is that it shows that
we should not always use the score problem to compute the winner
(as in dichotomous Young). The second is that by carefully looking at the
score problem, we may come up with easier winner problem
algorithms (as in Dodgson for single-peaked preferences).

\section{Preliminaries}

An election consists of a set of candidates $C$, and a set of voters $V$. Each voter
$v \in V$ has a corresponding vote (preference order) over the set of candidates.
The most commonly studied model is for voters to state their preferences as a total
order, i.e., strictly ranking all of the candidates from most to least preferred. Another
natural way for voters to state their preferences is as a dichotomous order, where
each voter approves of a subset of the candidates and disapproves of the remaining
candidates. More formally, given a set of candidates $C$, a dichotomous vote $v$ 
partitions $C$ into two sets, $A$ and $B$ (which may be empty), with the vote written as
$(A > B)$, such that for all $a \in A$ and for all $b \in B$, $v$ states $a > b$ (where $>$ denotes
strict preference) and $v$ does not state strict preference between candidates in $A$ nor between candidates in $B$.

An election system is a mapping from an election (a set of candidates and a set of
voters) to a subset of the candidate set referred to as the winner(s).
We are interested in the relationship between the complexity of the score and winner problems
for election systems so we focus on election systems with computationally
difficult score problems: Young~\cite{you:j:extending-condorcet}, Dodgson~\cite{dod:unpubMAYBE-without-embedded-citations:dodgson-voting-system}, and Kemeny~\cite{kem:j:no-numbers}, and some of their related
variants.

For our results and to define the election systems mentioned above, it will be useful
to refer to Condorcet winners and weak Condorcet winners.
A Condorcet winner is a candidate in an election that beats
each other candidate by pairwise majority.
Similarly, a weak Condorcet winner is a
candidate that beats-or-ties each other candidate by pairwise majority.

The Young score of a candidate is the size of a largest subset of voters
for which the candidate is
a weak Condorcet winner. A candidate is a Young winner if the candidate has
maximum Young score. Notice that the definition of Young applies to
dichotomous preferences.
We also consider the common slight variant of Young called strongYoung, in which
the goal is to make a candidate a Condorcet winner (instead of a weak Condorcet
winner). %

For total order votes, the Dodgson score of a candidate is the fewest number of swaps between adjacent
candidates in the voters' rankings such that the candidate can become a Condorcet winner.
A candidate is a Dodgson winner if the candidate has minimum Dodgson score. The
definition for weakDodgson is the same except we consider if a candidate can 
become a weak Condorcet winner.
For dichotomous preferences, a natural analogue for Dodgson is to
move candidates between the two groups in a dichotomous vote, which keeps the votes dichotomous, e.g., given the 
vote $(\{a,b\} > \{c,d\})$ we can move $c$ up to get $(\{a,b,c\} > \{d\})$ with one move. 

A total order $>$ is a Kemeny consensus if the sum of Kendall
tau distance to the voters is minimal, i.e.,
$\sum_{a > b} N(b,a)$ is minimal, where $N(b,a)$ denotes the number of voters that state $b > a$.
The Kemeny score
of a candidate $p$ is the minimal sum of Kendall tau distance to the voters for a total order 
that has $p$ as the most preferred.
For dichotomous preferences, we consider two variants
of Kemeny introduced by Zwicker~\shortcite{zwi:j:jk-kemeny} called
$(2,2)$-Kemeny and $(2,m)$-Kemeny.
In both systems, the
votes are dichotomous.\footnote{The definition of
dichotomous votes used by Zwicker~\shortcite{zwi:j:jk-kemeny} requires that each of the two
groups are nonempty. However, this does not make a difference in the results in our paper.}
In $(2,2)$-Kemeny, the consensus is also dichotomous.
In $(2,m)$-Kemeny, the consensus is a total order.
Having a total order consensus allows expressing more information about the electorate,
which sometimes may be more appropriate~\cite{ail:j:partial-rankings}.
In $(2,2)$-Kemeny and $(2,m)$-Kemeny,
the score of ranking $>$ is $\sum_{a > b} (N(a,b)-N(b,a))$ and we are looking to {\em maximize} the score.

The Slater rule is the qualitative version of the
Kemeny rule.
A total order $>$ is a Slater order
if it is closest to the pairwise
majority relation $>_m$ induced by the voters,
in the sense that we maximize the score, the number of ordered pairs of
candidates $(a,b)$ such that ($a > b$ if and only if $a >_m b$).
Slater winners are those candidates
that are ranked first in a Slater order.
The score of a candidate
$p$ is the maximal score of a Slater order 
that ranks $p$ first.

The definition of Slater order also naturally extends to weak orders.
A $k$-chotomous order $>$ is a
$k$-chotomous Slater order if it is
a $k$-chotomous order closest to $>_m$ in the
sense described above.
In $(2,k)$-Slater, the voters are dichotomous
and the Slater order is $k$-chotomous.
In $(2,m)$-Slater, the voters are dichotomous
and the Slater order is a total order.

We now define the score and winner decision problems with Young as an example. Note that
in Young we are looking to {\em maximize} the score (as is the case for $(2,2)$-Kemeny and $(2,m)$-Kemeny),
but in Dodgson and Kemeny we are looking
to {\em minimize} the score and so the decision problems for these systems must be adjusted accordingly.

\prob{YoungScore}{An election $(C,V)$, a candidate $p \in C$, and a number $k$.}{Is
the Young score of $p$ at least $k$?}

\prob{YoungWinner}{An election $(C,V)$ and a candidate $p \in C$.}{Does $p$
have highest Young score?}

Our computational results involve the classes \p, \np, and $\thetatwo$. The class
$\thetatwo$ was first studied by Papadimitriou and Zachos~\shortcite{pap-zac:c:two-remarks}, named by Wagner~\shortcite{wag:j:bounded}, and shown by
Hemachandra~\shortcite{hem:j:sky} to be equivalent to
$\p^{{\rm NP}}_{||}$, the class of problems solvable
by a polynomial-time oracle machine that
asks all of its queries to an NP oracle in parallel.
Note that ${\rm NP} \cup {\rm coNP} \subseteq \thetatwo
\subseteq \p^\np$.

\section{Dichotomous Preferences}\label{sec:res-d}

For total order votes, the winner problems for Young, strongYoung,
Dodgson, weakDodgson, and Kemeny are
$\thetatwo$-complete~\cite{bra-bri-hem-hem:j:sp2,rot-spa-vog:j:young,hem-hem-rot:j:dodgson,bra-bri-hem-hem:j:sp2,hem-spa-vog:j:kemeny}.
The $\thetatwo$ upper bounds for these problems are shown as follows: Use the 
associated NP-complete score problem, and compute
the scores of all candidates in parallel in polynomial time.

An election system is weakCondorcet-consistent if on every input that
has at least one weak Condorcet winner, the winners of the election system
are exactly the weak Condorcet winners.
If an election has dichotomous votes
then it has at least one weak Condorcet winner~\cite{ina:j:single-caved}, which
implies the following theorem.

\begin{theorem}\label{t:wcc}
The winner problem for a weakCondorcet-consistent election system
with dichotomous preferences is in \p. This holds even for election
systems that are weakCondorcet-consistent when restricted to
dichotomous preferences.
\end{theorem}

\subsection{Young Elections}

\begin{theorem}\label{t:ywd}
For dichotomous preferences,
YoungWinner  is in~\p\ and
YoungScore is \np-complete.
\end{theorem}

\begin{proofs}
Since Young is
weakCondorcet-consistent~\cite{fis:j:condorcet}, YoungWinner
in \p\ follows from Theorem~\ref{t:wcc}.

We will now show that YoungScore remains NP-complete for dichotomous preferences.
And so the ``natural'' way of deciding YoungWinner by using 
YoungScore as an oracle is not optimal.

We reduce from Independent-Set.
Given a graph $G = (V,E)$,
let the candidate set be $E \cup \{p\}$ and let the voter set consist of the following
voters.
\begin{itemize}
\item For each vertex $v \in V$, one voter corresponding to $v$ voting
 $(\{e \in E \ | \ v \in e \} > \cdots)$.
\item One voter voting $(\{p\} > \cdots)$.
\end{itemize}
Note that the Young score of $p$ is $\alpha(G) + 1$,
where $\alpha$ is the independence number of $G$, i.e., the size of a
maximum independent set of $G$. 
This score is realized
by the voter that ranks $p$ first and a set of voters corresponding to
a maximum-size independent set of $G$.~\end{proofs}

Note that the proof of the theorem above
does not contradict the previous statement that YoungWinner
is in P, since $p$ is clearly in general not a Young winner.
Also note that this construction
does not give NP-completeness for YoungScore for total orders.
Not surprisingly, it turns out that that problem is also
NP-complete (even to approximate~\cite{car-cov-fel-hom-kak-kar-pro-ros:j:dodgson}).
Somewhat surprisingly, a direct proof of NP-completeness is not given in the literature,
but it is implicit in the proof of Rothe, Spakowski, and Vogel~\shortcite{rot-spa-vog:j:young}.\footnote{It
is interesting to note that the NP-completeness of YoungScore does not directly follow
from the $\thetatwo$-completeness of YoungWinner and
strongYoungWinner: Under the
assumption that \np\ does not have p-measure 0,
there exists 
a set $A$ that is NP-complete under truth-table reductions,
but not NP-complete (under many-one reductions)~\cite{lut-may:j:cook-karp}.
Note that YoungWinner and strongYoungWinner are in
$\p_{tt}^A$, though $A$ is not NP-complete.}

It may be surprising that the construction for total orders is
harder than the one for dichotomous preferences, because it may seem
that a problem on total orders would be harder than the analogous problem 
for dichotomous preferences. However, for the plurality rule,
control by adding voters can be NP-complete for votes with ties~\cite{fit-hem:c:voting-with-ties}
while it is in P for total orders~\cite{bar-tov-tri:j:control}.
And it is easy
to see that dichotomous votes suffice to get
NP-completeness.

Add an additional vote $(\{p\} > \cdots)$ to get the following.

\begin{theorem}\label{t:sysd}
strongYoungScore for dichotomous preferences is \np-complete.
\end{theorem}

We now consider the ranking problem, which is the problem that
asks given two candidates $p$ and $r$ whether the score of $p$
is at least the score of $r$.
This will be used as an intermediate problem to show the
hardness of strongYoungWinner and for Young, to further contrast
the complexity of the score, ranking, and winner problems.
The construction also shows that the YoungLoser problem is
$\thetatwo$-complete, in stark contrast to the YoungWinner problem.\footnote{The
YoungLoser problem asks when given an election $(C,V)$ and a candidate $p \in C$,
if $p$ has lowest Young score.}

\begin{theorem}
\label{t:yrd}
YoungRanking for dichotomous preferences is $\thetatwo$-complete.
\end{theorem}

\begin{proofs}
Reduce from Min-Card-Independent-Set-Compare, in which we
are given two graphs $G$ and $H$ with the same number of vertices
and we ask if $\alpha(G) \geq \alpha(H)$. This problem
is $\thetatwo$-complete~\cite{wag:j:more-on-bh} (for
an explicit proof, see~\cite{spa-vog:c:theta-two-classic}).

Without loss of generality, assume
that $G$ and $H$ both contain at least one edge
(so that $\alpha(G) < \|V(H)\|$ and  $\alpha(H) < \|V(G)\|$)
and that the sets of vertices are
disjoint. Our reduction generalizes the YoungScore reduction
from the proof of Theorem~\ref{t:ywd}.
We will ensure that the Young score of $p$ is $\alpha(G) + 1 + \|V(H)\| + 1$
and that the Young score of $r$ is $\alpha(H) + 1 + \|V(G)\| + 1$.
This proves the theorem, since $\|V(G)\| = \|V(H)\|$.

Let the candidate set be $E(G) \cup E(H) \cup \{p,r\}$ and let the
voter set consist of the following voters.
\begin{description}
    \item[Type~I] For each $v \in V(G)$, one voter corresponding to $v$ voting
$(\{r\} \cup E(H) \cup \{e \in E(G) \ | \ v \in e \} > \cdots)$.

    \item[Type~II] One voter voting $(E(H) \cup \{p,r\}  > \cdots)$.

    \item[Type~III] For each $v \in V(H)$, one voter corresponding to $v$ voting
$(\{p\} \cup E(G) \cup \{e \in E(H) \ | \ v \in e \} > \cdots)$.

    \item[Type~IV] One voter voting $(E(G) \cup \{p,r\}  > \cdots)$.
\end{description}
Note that to realize the Young score of $p$, we should always include
all the Type~III and Type~IV voters. The rest of the argument is as in
the proof of Theorem~\ref{t:ywd}. Note that $p$ ties-or-beats each 
candidate in $E(H)$, since we are including $\alpha(G) < \|V(H)\|$
Type I votes. 
\end{proofs}

Add a vote $(E(H) \cup \{p,r\}  > \cdots)$
and a vote $(E(G) \cup \{p,r\}  > \cdots)$ to get the following.

\begin{theorem}\label{t:syrd}
strongYoungRanking for dichotomous preferences is $\thetatwo$-complete.
\end{theorem}

As stated in Theorem~\ref{t:ywd}, YoungWinner is in \p\ for dichotomous votes.
In contrast, the winner problem for strongYoung for dichotomous votes is
$\thetatwo$-complete.

\begin{theorem}\label{t:sywd}
strongYoungWinner for dichotomous preferences is $\thetatwo$-complete.
\end{theorem}

\begin{proofs}
The main insight here is that we
can always make sure that a 
candidate $c$'s strongYoung score is 0, by adding a candidate $c'$ and
making sure that $c$ and $c'$ are tied in every vote.
We adapt the construction used to show that strongYoungRanking
for dichotomous preferences is $\thetatwo$-complete from Theorem~\ref{t:syrd}.
For each $e \in E(G) \cup E(H)$,
add a candidate $e'$ and make sure that $e$ and $e'$ are tied in 
every vote.~\end{proofs}

It is interesting to see that our $\thetatwo$-completeness proofs
for dichotomous preferences are also significantly easier than those
for total orders~\cite{rot-spa-vog:j:young}.

Note that the approach from Theorem~\ref{t:sywd} does not work for YoungWinner.
Rothe, Spakowski, and Vogel~\shortcite{rot-spa-vog:j:young} reduce the
strongYoungRanking to
strongYoungWinner problem by replacing each candidate $g$ other than
$c$ and $d$ by $\|V\|$ new candidates $g_0, \ldots, g_{\|V\|-1}$,
and by replacing the occurrence of $g$ in the $i$th voter by
$g_i > g_{i+1} > \cdots > g_{i + \|V\| - 1}$ (modulo $\|V\|$).
This does not change the scores of $c$ and $d$, but ensures that
the strongYoung score of every other candidate is at most 1, and so
we can ensure that these candidates are never winners in the
image of the reduction.
Note that this construction does not work for dichotomous preferences
(which is consistent with the Theorem~\ref{t:ywd} result that
YoungWinner for dichotomous preferences is in P).
The construction also does not work for trichotomous preferences,
or indeed for any $k$-chotomous preferences.

However, we can adapt the construction used to show that YoungRanking
for dichotomous preferences is $\thetatwo$-complete from Theorem~\ref{t:yrd}.

\begin{theorem}\label{t:ywt}
YoungWinner for trichotomous preferences is $\thetatwo$-complete.
\end{theorem}

\begin{proofs}
The main insight is that we
can make sure that 
candidate $c$'s Young score is relatively low, by adding candidates
$c'$ and $c''$, tied with $c$ in each original vote,
and adding, for some big integer $B$, the following $6B$ voters.
\begin{itemize}
\item $B$ voters voting $(c > c' > \cdots)$.
\item $B$ voters voting $(\cdots > c > c')$.
\item $B$ voters voting $(c' > c'' > \cdots)$.
\item $B$ voters voting $(\cdots > c' > c'')$.
\item $B$ voters voting $(c'' > c > \cdots)$.
\item $B$ voters voting $(\cdots > c'' > c)$.
\end{itemize}
Note that
the Young scores of $c$, $c'$, and $c''$ are at most
$n - 2B$, where $n$ is the number of voters. %

We adapt the construction used to show that YoungRanking is
$\thetatwo$-hard from Theorem~\ref{t:yrd}.
For each $e \in E(G) \cup E(H)$, we add $e'$ and $e''$,
we replace each occurrence of $e$ in the original votes by $e,e',e''$,
and we add $6\|V(G)\|$ voters as specified above, taking $B = \|V(G)\|$.
For readability, we write
$\widehat{E}$ for $\{e,e',e'' \ | \ e \in E\}$.
This gives the following set of voters.

\begin{description}
    \item[Type~I] For each vertex $v \in V(G)$, 
    \begin{itemize}
  \item One voter corresponding to $v$ voting\\
      $(\{r\} \cup \widehat{E(H)} \cup \widehat{{\{e \in E(G) \ | \ v \in e \}}} > \cdots)$.
    \end{itemize}
    \item[Type~II] \hfill
    \begin{itemize}
\item One voter voting $(\widehat{E(H)} \cup \{p,r\}  > \cdots)$.
    \end{itemize}
    \item[Type~III] For each vertex $v \in V(H)$, 
    \begin{itemize}
  \item One voter corresponding to $v$ voting\\
$(\{p\} \cup \widehat{E(G)} \cup  \widehat{\{e \in E(H) \ | \  v \in e \}} > \cdots)$.
    \end{itemize}
    \item[Type~IV] \hfill
    \begin{itemize}
\item One voter voting $(\widehat{E(G)} \cup \{p,r\}  > \cdots)$.
    \end{itemize}
    \item[Type~V] For each $e \in V(G) \cup V(H)$, 
    \begin{itemize}
\item $\|V(G)\|$ voters voting $(e > e' > \cdots)$.
\item $\|V(G)\|$ voters voting $(\cdots > e > e')$.
\item $\|V(G)\|$ voters voting $(e' > e'' > \cdots)$.
\item $\|V(G)\|$ voters voting $(\cdots > e' > e'')$.
\item $\|V(G)\|$ voters voting $(e'' > e > \cdots)$.
\item $\|V(G)\|$ voters voting $(\cdots > e'' > e)$.
    \end{itemize}
\end{description}
It is immediate that for all $c \neq p,r$, 
$\ys(c) \leq n - 2\|V(G)\|$.
We can show that
$\ys(p) = n - \|V(G)\| + \alpha(G)$ and
$\ys(r) = n - \|V(H)\| + \alpha(H)$.
Since $\|V(G)\| = \|V(H)\|$, it follows that
$\alpha(G) \geq \alpha(H)$ if and only if $p$ is a Young winner.

Let $W$ be a set of voters that realizes the Young score of $p$
with a maximal number of Type V voters.
Note that all the voters that rank $p$ tied for first are in $W$.
So, all Type II, III, and IV voters, and half of the Type V voters
are in $W$.
 
Next, we will show that $W$ contains all Type V voters.
Suppose it does not. Then there is an $e \in E(G)$ such that
there is a voter $w$
voting  $e > e' > \cdots$ or $e' > e'' > \cdots$ or $e'' > e > \cdots$
that is not in $W$. Since $W$ contains a maximal
number of Type V voters, adding $w$
to $W$ will cause $p$ to not be a weak Condorcet winner.
It follows that $W$ contains a Type I voter $w'$ that prefers
$e$, $e'$, and $e''$ to $p$. 
But note that $p$ is still a weak Condorcet winner in $W$ if
we replace $w'$ by $w$ (since $p$ does not do worse against any candidate
in $w$ compared to $w'$). This contradicts the assumption
that $W$ contains a maximal number of Type V voters.
 
So, $W$ contains all Type II, III, IV, and V voters.
Then $W$ contains $\alpha(G)$ Type I voters corresponding to a maximum 
independent set of $G$ (recall that $\alpha(G) <  \|V(H)\|$).
\end{proofs}

\subsection{Dodgson Elections} %

We now show that the complexity behavior of Young for dichotomous preferences
does not occur for Dodgson and Kemeny for dichotomous
preferences.

Recall that for Dodgson for dichotomous preferences a candidate can only be swapped
up from disapproved to approved or swapped down from approved to disapproved.
So unlike in the case for total orders, to determine the score of a candidate we
cannot just consider swaps that move that candidate up. For example, given the
vote $(\{a,b\} > \{c, d\})$, for $c$ to beat $a$ pairwise, we first need to move $c$
up to get $(\{a,b,c\} > \{d\})$ and then move $a$ down to get $(\{b,c\} > \{a,d\})$.
(We
could of course also do these moves in the
reverse order.) With this in mind, we can show that DodgsonScore and weakDodgsonScore are each in \p\ (and so are
the corresponding winner problems).

\begin{theorem}\label{t:dsd}
DodgsonScore and weakDodgsonScore, for dichotomous preferences, are each in~\p.
\end{theorem}

\begin{proofs}
First notice that moving $p$ up from the set of disapproved candidates in a vote
to the set of approved candidates decreases 
$N(a,p) - N(p,a)$
by 1 for
every candidate $a \in C-\{p\}$. And it is not possible to decrease
$N(a,p) - N(p,a)$
by more than 1 with one move.
So, the weakDodgson score of $p$ is the max over all candidates $a \in C-\{p\}$ of
$N(a,p) - N(p,a)$:
We can make $p$ a weak Condorcet winner by moving
$p$ up from disapproved to approved in that many votes and it is easy to see that
there are enough votes to do this.
So, weakDodgsonScore is in~\p.

For DodgsonScore, if $p$ starts out as a Condorcet winner, the score is 0.
Otherwise, we need to move $p$ up in one more vote compared to what was
needed to make $p$ a weak Condorcet winner. If no such vote exists then for
some candidate $a$, $a > p$ in each original vote. In this case, move $p$ up
in every vote, and then for every candidate $a$ for which $a$ and $p$ are tied,
move that candidate down. So, DodgsonScore is also in \p. 
\end{proofs}

\subsection{Kemeny Elections}

Zwicker~\shortcite{zwi:j:jk-kemeny} shows that the winner problems
for $(2,2)$-Kemeny and $(2,m)$-Kemeny are in P. 
An easy way to see this for $(2,2)$-Kemeny is because this is the
same as the mean rule~\cite{zwi:j:jk-kemeny}.
Computing the score of a candidate $p$
is a little harder, since we need to rank $p$ (tied for) first in a dichotomous ranking, and so 
we are not ranking according to approval score (as in the mean rule).
However, 
a similar argument to what is used in the proof of Lemma~3 from
Zwicker~\shortcite{zwi:j:jk-kemeny}, which computes a
$k$-chotomous consensus, can be used to show that the
score problem for $(2,2)$-Kemeny is also in \p.

For Kemeny versions that require a total order consensus
(such as $(2,m)$-Kemeny)
we will show that the score problem polynomial-time Turing-reduces to
the winner problem, which implies that the score problem
is in P if the winner problem is in P. And it follows
that $(2,m)$-KemenyScore is in P.

\begin{theorem}\label{thm:kemeny-score-to-winner}
For each preference domain ${\cal D}$ that is closed
under deletion of candidates,
KemenyScore for ${\cal D}$-preferences polynomial-time Turing-reduces to
KemenyWinner for ${\cal D}$-preferences with a total order consensus.
\end{theorem}

\begin{proofs}
The reduction in the theorem above works in the
following way. To compute the Kemeny score of candidate $p$,
we need to compute the score of a total order that ranks
$p$ first. So, put $p$ first in the total order.
The contribution of $p$ to the score is independent of how the 
remaining candidates are ordered (for example, $p$ contributes
$\sum_{c \neq p} (N(p,c) - N(c,p))$ for $(2,m)$-Kemeny).
For an optimal order, we need to order 
$C - \{p\}$ such that the order restricted to those
candidates is optimal, i.e., we need to compute a
Kemeny consensus of the electorate restricted to $C - \{p\}$.

So, delete $p$ and repeatedly query whether a candidate $a$ is a winner.
If so, put $a$ next in the order and delete $a$. This
builds an optimal total order with $p$ first.
It is clear this argument
holds for preference domains that are
closed under deletion of candidates and this includes
total order, dichotomous, %
single-peaked, and single-crossing preferences.~\end{proofs}
\subsection{Slater Elections}
For total-order votes, the score and winner problems for Slater are each
\np-hard~\cite{bar-tov-tri:j:who-won}.

We first mention that the analogous result to Theorem~\ref{thm:kemeny-score-to-winner} holds for Slater,
with the proof using essentially the same argument.

\begin{theorem}\label{thm:slater-score-to-winner}
For each preference domain ${\cal D}$ that is closed
under deletion of candidates,
SlaterScore for ${\cal D}$-preferences polynomial-time Turing-reduces to
SlaterWinner for ${\cal D}$-preferences with a total order consensus.
\end{theorem}

We now consider the complexity of the score and winner problems for Slater
for dichotomous votes.

\begin{theorem}
For all fixed $k$, the winner and score problems
for $(2,k)$-Slater and $(2,m)$-Slater are in P.
\end{theorem}

\begin{proofs}
Since Slater is weakCondorcet-consistent,
it follows from Theorem~\ref{t:wcc} that the winner problem
for $(2,m)$-Slater is in \p, and thus Theorem~\ref{thm:slater-score-to-winner}
implies that $(2,m)$-SlaterScore is in \p.

It remains to show that $(2,k)$-SlaterScore is in \p.
We know from Inada~\shortcite{ina:j:single-caved} that the induced majority relation
$>_m$ is a weak order.
Below, we will prove the following lemma.

\begin{lemma}
\label{l:slater-order}
Let $>_m$ be a weak order majority relation and let
$>$ be a $k$-chotomous weak order. If there exist
two candidates $a,b$ such that $a >_m b$ and $b > a$,
then switching $a$ and $b$ in $>$ will give a
$k$-chotomous weak order with higher Slater score.
\end{lemma}

From this lemma, it follows that 
the Slater score of candidate $p$ is
witnessed by an order $>$ such that $p$ is a top-ranked candidate
and such that for all $a,b \neq p$,
if $a >_m b$, then not $b > a$.  What this means is that,
apart from candidate $p$, the order witnessing the Slater
score is not inconsistent with the majority relation.

Since $k$ is constant, there are polynomially many
sequences of positive integers $\ell_1, \ldots, \ell_k$ that add to $m$.
View such a sequence as a $k$-chotomous order 
not inconsistent with $>_m$ that ranks $p$ first.
Note that this order is not unique, but the Slater score
of such an order is unique. This gives a polynomial-time
algorithm to compute the Slater score of~$p$.~\end{proofs}

It remains to prove Lemma~\ref{l:slater-order}.

\begin{proofs}
It is immediate that switching $a$ and $b$ gives a
$k$-chotomous order. It remains to show that the score
of this order, which we will call $>'$, is higher than
that of the original order $>$.

First note that $\{a,b\}$ contributes 0 to the Slater score
of $>$ and 2 to the Slater score of $>'$,
since $a >_m b$ if and only if $a >' b$ and
$b >_m a$ if and only if $b >' a$.
For $c,d \not \in \{a,b\}$, the contribution of
$\{c,d\}$ to the Slater score of $>$ and $>'$ are the same.
We will now show that for every $c \not \in \{a,b\}$,
the contribution of $\{a,c\}$ plus the contribution
of $\{b,c\}$ to the Slater score of $>'$ is never less
than the contribution to the Slater score of $>$.

This proceeds by a simple case distinction shown in Table~\ref{tbl:slater}. We have five possibilities
for $>_m$ and five possibilities for $>$.
We handle each of the 25 cases separately for clarity.~\end{proofs}

\begin{table*}
\centering
\small
\begin{tabular}{l|c|c|c|c|c|}
& $c > b > a$ & $\{b,c\} > a$ & $b > c > a$ & $b > \{a,c\}$ & $b > a > c$\\
& $c >' a >' b$ & $\{a,c\} >' b$ & $a >' c >' b$ & $a >' \{b,c\}$ & $a >' b >' c$\\
\hline
$c >_m a >_m b$ &2+2; 2+2 &2+1; 1+2 &2+0; 0+2 &1+0; 0+1 &0+0; 0+0 \\  
$\{a,c\} >_m b$ &1+2; 1+2 &1+1; 2+2 &1+0; 1+2 &2+0; 1+1 &1+0; 1+0   \\
$a >_m c >_m b$ &0+2; 0+2 &0+1; 1+2 &0+0; 2+2 &1+0; 2+1 &2+0; 2+0\\
$a >_m \{b,c\}$ &0+1; 0+1 &0+2; 1+1 &0+1; 2+1 &1+1; 2+2 &2+1; 2+1\\
$a >_m b >_m c$ &0+1; 0+0 &0+1; 1+0 &0+2; 2+0 &1+2; 2+1 &2+2; 2+2
\end{tabular}
\caption{Case distinctions for the proof of Lemma~\ref{l:slater-order}.
The content
of each cell is the $\{a,c\}$ contribution for $>$ ``+''
the $\{b,c\}$ contribution for $>$, and %
the $\{a,c\}$ contribution for $>'$ ``+''
the $\{b,c\}$ contribution for $>'$.}
\label{tbl:slater}
\end{table*}

\section{Single-Peaked Preferences}\label{sec:res-sp}

Single-peaked preferences model the preferences of the electorate
with respect to a one-dimensional axis $L$, a total ordering of the
candidates, where each voter has a single most preferred candidate
(their peak) and candidates farther to the leftmost (rightmost) ends
of the axis are strictly less preferred. More formally, for every triple of
candidates $a L b L c$ or $c L b L a$, for
every voter $v \in V$ if $v$ states $a > b$ then $v$ states $b > c$.

A real-life scenario for single-peaked preferences is voting for the
temperature for a room. When presented with several options (candidates) for
the temperature, the preferences of the voters can be explained by a
left-to-right ordering of the options from the lowest to the highest,
where each voter has a most-preferred temperature, and has strictly
decreasing preferences for lower temperatures and has strictly decreasing
preferences for higher temperatures.

The pairwise majority relation for single-peaked preferences is transitive.
Brandt et al.~\shortcite{bra-bri-hem-hem:j:sp2} show that it follows
from their~Theorem~3.2 (the analogue of Theorem~\ref{t:wcc})
that the winner problems for Kemeny, Young, and weakDodgson are in P.
They also show that the winner problems for
strong\-Young and Dodgson are in P.

So, what
happens to the single-peaked score problems of the systems mentioned above?

\subsection{Dodgson Elections}
Peters~\shortcite{pet:c:sptu}
states that
``... while Brandt et al. (2015) give an efficient algorithm for finding a
Dodgson \emph{winner} in the case of single-peaked preferences, the problem of efficiently
calculating \emph{scores} appears to be open and non-trivial.''
The reason for this nontriviality is that after swapping, 
the electorate is not required to be single-peaked
(see~\cite[Footnote 5]{bra-bri-hem-hem:j:sp2}).
This makes single-peaked Dodgson very different
from single-peaked Kemeny and Young, where we will never get
non-single-peaked electorates in the computation. We will show below
that for Young and Kemeny, the single-peaked score problem is in~P.

More surprisingly, we also show the following theorem. 

\begin{theorem}\label{thm:ds}
DodgsonScore and weakDodgsonScore for single-peaked preferences are in \p.
\end{theorem}

\begin{proofs}
We start with an example that shows the core argument of our algorithm.
Suppose the societal axis is $a_1 L a_2 L a_3 L a_4 L p$
and suppose our election
consists of the following types of votes.

10 votes of the form $(\{a_1, a_2, a_3, a_4\} > p)$.\footnote{In contrast to the results
in the previous section, this result is about total-order
votes. Here we are using sets to denote those candidates strictly ranked in that location of a total-order vote, e.g., a vote of the form $(\{x,y\} > z)$ corresponds to the total-order votes
$(x > y > z)$ and $(y > x > z)$.}

50 votes of the form $(\{a_2, a_3, a_4\} > p > \cdots)$.

10 votes of the form $(\{a_3, a_4\} > p > \cdots)$.

20 votes of the form $(a_4 > p > \cdots)$.

11 votes of the form $(p > \cdots)$.

\noindent
In order for $p$ to become a Condorcet winner,
every candidate other than $p$ can be preferred to $p$ by at
most 50 voters.
$p$ does not need any votes over $a_1$, and we leave
the 10 votes of the form $(\{a_1, a_2, a_3, a_4\} > p)$ as is.
$p$ needs 10 votes over $a_2$. In 10 of the votes of the form
$(\{a_2, a_3, a_4\} > p > \cdots)$, we swap $p$ to the top
of the order (using 30 swaps total) and we
leave the remaining 40 votes of this form as is.
Note that in the resulting  set of votes:

10 votes of the form $(\{a_1, a_2, a_3, a_4\} > p)$.

40 votes of the form $(\{a_2, a_3, a_4\} > p > \cdots)$.

10 votes of the form $(p > \{a_2, a_3, a_4\} >  \cdots)$.

\noindent
$a_2$, $a_3$, and $a_4$ are preferred to $p$ by 50 voters.
This implies that we need to swap
$p$ to the top of the preference order in all 
votes of the form $(\{a_3, a_4\} > p > \cdots)$ and all
votes of the form $(a_4 > p > \cdots)$.
Note that
no swap is wasted, and so $p$ has a
Dodgson score of~70.

We now show that this approach can be generalized to all single-peaked
electorates.
Let the societal axis be $a_1 L a_2 L \dots L a_{m_a} L p L
b_{m_b} L \dots L b_2 L b_1$. Our goal is to compute the (Dodgson
or weakDodgson) score of $p$.
Let $H$ (for ``half'') be the maximum number of voters that
can prefer $c$ to $p$ while making $p$ a winner.
To be precise, in the case of Dodgson winner,
$H$ is $\lfloor (n+1)/2 \rfloor$, and in the case of 
weakDodgson winner, $H$ is $\lfloor n/2 \rfloor$, where
$n$ is the number of voters. In the example above, $H$ is 50.
We will show that we can make $p$
a winner without wasting swaps, i.e., in the election
that makes $p$ a winner,
$c$ is preferred to $p$ by exactly $H$ voters for
every candidate $c$ such that $N(c,p) > H$, and
by $N(c,p)$ voters
for every candidate $c$ such that $N(c,p) \leq H$.
Since we are not wasting swaps, this witnesses
the Dodgson score of $p$ (with value $\sum_{c \neq p, N(c,p) > H} (N(c,p) - H)$).

Note that in each single-peaked vote, $p$ is preferred to
all $a$ candidates or to all $b$ candidates. So, when we swap $p$
up in a vote, this will help $p$ against $a$ candidates or
against $b$ candidates, but not against both. This implies that we can
treat the $a$ candidates and $b$ candidates separately. We
will show that we can ensure that each $a$ candidate is preferred
to $p$ by at most $H$ voters
without wasting swaps.
The same holds for the $b$ candidates, which proves the theorem.

As in the example, we classify the voters in terms of the
set of $a$ candidates that are preferred to $p$. For every $i, 1 \leq i
\leq m_a$, let $k_i$ be the number of voters with a vote of the form
$(\{a_i, a_{i+1}, \ldots, a_{m_a}\} > p > \cdots)$ and let
$k_0$ be the number of voters that prefer $p$ to all $a$ candidates.
So, $N(a_i,p) = \sum_{j = 1}^i k_j$, the total number of voters
is $n = \sum_{i=0}^{m_a} k_i$, and $H$ is defined as above.

If for all $1 \leq i \leq m_a$, $N(a_i,p) \leq H$, then
$p$ does not need to gain any points over $a$ candidates.
Otherwise, let $i_0$ be the smallest index such that
$N(a_{i_0},p) > H$.
We now swap $p$ to top of the preference order of 
$N(a_{i_0},p) - H$ voters with a vote of the form 
$(\{a_{i_0}, a_{i_0+1}, \ldots, a_{m_a}\} > p > \cdots)$.\footnote{Note that
$N(a_{i_0},p) - H \leq k_{i_0}$. This is immediate
if $i_0 = 1$. If $i_0 > 1$, $N(a_{i_0-1},p) \leq H$, and
thus $N(a_{i_0},p) \leq H + k_{i_0}$.}
Also, for all $i, i_0 < i \leq m_a$,
we swap $p$ to the top of the preference order
of all the voters with a vote of the
form $(\{a_{i}, a_{i+1}, \ldots, a_{m_a}\} > p > \cdots)$.

We now show that after swapping, 
each $a$ candidate is preferred to $p$ by at most $H$ voters
without wasting swaps. 
That is,
$a_i$ is preferred to $p$ by $H$ voters for
all $i \geq i_0$ and
$a_i$ is preferred to $p$ by $N(a_i,p)$ voters
for all $i < i_0$.

For $i < i_0$, since we never swap $p$ over $a_i$, it is 
immediate that $a_i$ is preferred to $p$ by the same voters
as in the original election.
For $i \geq i_0$, 
$a_i$ is preferred to $p$ by all
unchanged voters that prefer $a_i$ to $p$.
And there are exactly $\sum_{i = 1}^{i_0 - 1}k_i + k_{i_0} - 
(N(a_{i_0},p) - H) = N(a_{i_0},p) - (N(a_{i_0},p) - H) =
H$ such voters.
\end{proofs}

The proof of the above theorem shows that for single-peaked
electorates, the Dodgson score of $p$ is equal to
$\sum_{c \neq p, N(c,p) > H} (N(c,p) - H)$.
This also gives a simple algorithm for DodgsonWinner.
Though this problem was known to be in P~\cite{bra-bri-hem-hem:j:sp2},
that algorithm was more complicated, since it did not look at the form
of the entire single-peaked electorate in the way we do in our proof.

\subsection{Young Elections}\label{sec:young-sp}
It is easy to see that we can adapt the construction for Dodgson
from the previous section to Young.
For every voter that we swap $p$ to the top of the preference order
of in the Dodgson construction, we now delete that voter.
It is easy to see that this
gives the minimum number of voters to delete in order for
$p$ to become a Condorcet (weak Condorcet) winner, and
so the number of remaining voters is exactly the strongYoung (Young) score
of $p$.
This result also follows from the result that constructive control by
deleting voters for Condorcet and weak Condorcet elections are each in \p\ for single-peaked
preferences~\cite{bra-bri-hem-hem:j:sp2}.

\subsection{Kemeny Elections}
Brandt et al.~\shortcite{bra-bri-hem-hem:j:sp2} show that KemenyWinner for single-peaked
preferences is in P.
Since we are computing a total order consensus, it
follows from Theorem~\ref{thm:kemeny-score-to-winner} that
KemenyScore for single-peaked preferences Turing-reduces to KemenyWinner
for single-peaked preferences, and
is thus also in P.

\subsection{Slater Elections}
For single-peaked preferences, computing a Slater winner is in \p\ (see Conitzer~\shortcite{con:c:slater}).
As with Kemeny, since we are computing a total order consensus, it
follows from Theorem~\ref{thm:slater-score-to-winner} that
SlaterScore for single-peaked preferences Turing-reduces to SlaterWinner
for single-peaked preferences, and
is thus also in \p.

\section{Single-Crossing Preferences}

Another important domain restriction that ensures the majority relation
is transitive is the single-crossing restriction~\cite{mir:j:single-crossing},
where the voters can be ordered along a one-dimensional axis $L =  v_1 L v_2 L \ldots L v_n$
such that for each pair of candidates $a,b \in C$ all of the voters that state $a > b$
precede the voters that state $b > a$, i.e., there is a single crossing
point for each pair.

A real-life scenario for single-crossing preferences is the traditional liberal-conservative
political spectrum, for the case where voters' preferences depend only on
where they fall on the political spectrum. The voters can be ordered along an
axis where the leftmost (rightmost) voters are the most-liberal
(most-conservative), and for each choice between two alternatives,
the preferences of the voters swaps at most once when moving left-to-right
along the axis. If the alternatives can also be placed on this spectrum
(this happens for example when the alternatives correspond to political
candidates), the preferences will also be single-peaked.\footnote{Not all
single-crossing preferences are single-peaked (see Example~\ref{ex:sc}).
Single-peaked preferences are also not necessarily single-crossing.
For example, in the voting for the temperature in a room example, the
(single-peaked) votes $(16 > 18 > 21 > 25)$, $(18 > 21 > 25 > 16)$, and
$(21 > 18 > 16 > 25)$ are not single-crossing.}

As in the case for dichotomous preferences and for single-peaked
preferences, we immediately obtain that YoungWinner and weakDodgsonWinner
are each in P.
It follows from Magiera and Faliszewski~\shortcite{mag-fal:j:single-crossing-control}
that
strongYoungScore and strongYoungWinner are each in \p.
As mentioned by Cornaz et al.~\shortcite{cor-gal-spa:c:kemeny-spsc-width},
KemenyWinner for single-crossing elections is in \p. And due to the weakCondorcet-consistency
of Slater, SlaterWinner for single-crossing elections is also in \p.
This immediately implies using Theorem~\ref{thm:kemeny-score-to-winner}
that KemenyScore is in \p, and using Theorem~\ref{thm:slater-score-to-winner} that
SlaterScore is in \p.

An important property for the complexity is the fact that 
single-crossing elections have a variant of the median voter 
theorem, in that the median voter(s)
represent the majority relation~\cite{rot:j:representative-voter,gan-sma:j:single-crossing}.
The following corollary extracts the properties that we need.

\begin{corollary}\label{cor:sc}
Let $v_1 L v_2 L \dots L v_n$ be a single-crossing order of voters.
A candidate $p$ is a weak Condorcet winner if and only if
\begin{itemize}
\item $n$ is odd and $p$ is the most preferred candidate of the median 
voter.
\item $n$ is even and for every $a \neq p$, $a$ is preferred to $p$ by at
most one of the two median voters (and $a > p$ by
exactly one of the median voters if and only if $p$ and $a$ are tied). 
\end{itemize}
\end{corollary}

\begin{theorem}\label{thm:youngscore-sc}
YoungScore for single-crossing preferences is in \p.
\end{theorem}
\begin{proofs}
For every voter $v$ that has $p$ at the top of its preference order, 
keep a maximum odd number of voters such that $v$ is the median voter.
If the number of remaining voters is greater than the current best score,
this becomes the current best score.

For every pair of voters $v$ and $w$, if there is a candidate
$a \neq p$ that is preferred to $p$ by $v$ and $w$, go to the next
loop iteration. Otherwise, keep a maximum even number of voters
such that $v$ and $w$ are the two median voters.
If the number of remaining voters is greater than the current best score,
this becomes the current best score.
\end{proofs}

\begin{theorem}
DodgsonWinner for single-crossing preferences is in~\p.
\end{theorem}

\begin{proofs}
This is trivial if there are an odd number of voters. So, assume that
the number of voters is even.
We will show that every Dodgson winner is a weak Condorcet winner,
and that the Dodgson scores of weak Condorcet winners are easy to compute.
This immediately implies the theorem, since the Dodgson winners are the
weak Condorcet winners with lowest Dodgson score.

Suppose $p$ is a weak Condorcet winner.
It follows from Corollary~\ref{cor:sc} that
for every candidate $a \neq p$,
$p$ is preferred to $a$ by at least one of the median voters and 
$a$ is preferred to $p$ by one of the median voters if and only
if $a$ is tied with $p$ pairwise. To make $p$ a Condorcet winner
with a minimal number of swaps, it suffices to swap $p$ to the top
of the two median voters. This gives a Dodgson score
of $\|\{a \in C - \{p\} \ | \ p \mbox{ ties } a\}\|$. 

If $q$ is not a weak Condorcet winner, then there is a weak Condorcet 
winner $p$ such that $p$ beats $q$ pairwise. It is easy to see that 
the Dodgson score of $q$ is greater than the Dodgson score of $p$,
since for every candidate $a \in C - \{p,q\}$,
if $p$ needs a vote over $a$, then so does $q$. In addition, $q$ needs
two votes over $p$. It follows that the Dodgson score of $q$
is greater than
$\|\{a \in C - \{p\}\ | \ p \mbox{ ties } a\}\|$ which is the Dodgson score of $p$.
\end{proofs}

It is interesting to see that the algorithm for DodgsonWinner for single-crossing
preferences from the previous theorem
is similar to the algorithm  for single-peaked preferences
from Brandt et al.~\shortcite{bra-bri-hem-hem:j:sp2}, in that there it also
was shown that only
weak Condorcet winners can be winners and how to compute the
Dodgson score of a weak Condorcet winner.
In Theorem~\ref{thm:ds}, 
we finally solved the open problem of
computing the Dodgson and weakDodgson scores for single-peaked preferences.
We have not managed to solve the complexity of these problems for
single-crossing electorates. Recall that it was crucial that
in the single-peaked case we could always realize the score without
wasting swaps. That is not the case in the single-crossing case, as
shown by the following simple example. This gives some indication
that the single-crossing case may be harder to handle.

\begin{example}\label{ex:sc}\normalfont
Consider the following four voters, single-crossing w.r.t.\
the ordering $v_1 L v_2 L v_3 L v_4$.

\begin{itemize}
\item $v_1$ voting $(a > b > p > c)$.

\item $v_2$ voting $(a > b > p > c)$.

\item $v_3$ voting $(a > c > p > b)$.

\item $v_4$ voting $(a > c > p > b)$.
\end{itemize}

\noindent
To become a Condorcet winner $p$
needs one vote over $b$, one vote over $c$, and three votes
over $a$. However, the three votes over $a$ can only be obtained
by wasting an extra swap over either $b$ or $c$. So the Dodgson
score of $p$ is six.
\end{example}

\section{Future Work}
A concrete open question is the complexity of Dodgson score for
single-crossing elections. Using the phrasing of Peters~\shortcite{pet:c:sptu}:
While we give an efficient algorithm for finding a
Dodgson \emph{winner} in the case of single-\emph{crossing} preferences,
the problem of efficiently
calculating \emph{scores} appears to be open and nontrivial.

We also point out that there
are other options for ``dichotomous Kemeny'' depending on
the amount a tie contributes to the distance.
Kemeny~\shortcite{kem:j:no-numbers} adds 0.5
and Fagin et al.~\shortcite{fag-kum-mah:j:comparing-partial-rankings}
considers all penalty values between 0 and 1.

It should be noted that $(2,2)$-Kemeny does not correspond to
any of these penalty values, since this system depends solely on
the induced weighted majority graph (making it a C2 rule in the sense of Fishburn~\shortcite{fis:j:condorcet}),
whereas the others do not.
Computing a dichotomous consensus for $p > 0$
is a very interesting challenge.
(For the $p = 0$ case, this can be solved in polynomial time via
network flow. Computing a total order consensus is equivalent to $(2,m)$-Kemeny
and thus also in \p.)

\bigskip

\noindent
{\bf Acknowledgements:}
This work was supported in part by
NSF-DUE-1819546, a Renewed Research Stay grant from the
Alexander von Humboldt Foundation, and NSF
Graduate Research Fellowship
DGE-1102937.
Research done in part while Zack Fitzsimmons was on
research leave at Rensselaer Polytechnic Institute
and in part while Edith
Hemaspaandra was on sabbatical visits to 
ETH-Z{\"u}rich and Heinrich-Heine-Universit{\"a}t D{\"u}sseldorf.
We thank Lane Hemaspaandra, Dominik Peters, and reviewers for helpful comments.

\end{document}